\documentclass[10pt,journal,compsoc]{IEEEtran}
\usepackage{graphicx}
\usepackage{amsmath}
\usepackage{subcaption}
\usepackage{comment}
\usepackage{amsmath}
\usepackage[english]{babel}
\usepackage{inputenc}
\usepackage[linesnumbered,algoruled,boxed,lined]{algorithm2e}
\usepackage{amsfonts}
\usepackage{amssymb}

\ifCLASSOPTIONcompsoc
  \usepackage[nocompress]{cite}
\else
  \usepackage{cite}
\fi
\ifCLASSINFOpdf
\else
\fi
\begin{document}
%
\title{A Two-Phase Dynamic Throughput Optimization Model for Big Data Transfers}
%
%
%
%

\author{MD~S~Q~Zulkar~Nine,
        and~Tevfik~Kosar,~\IEEEmembership{Member,~IEEE}

\IEEEcompsocitemizethanks{\IEEEcompsocthanksitem This is a preliminary draft submitted to ARXIV.COM. Please note that the actual published version of the paper may be different than this version.}
}

\IEEEtitleabstractindextext{%
\begin{abstract}
The amount of data moved over dedicated and non-dedicated network links increases much faster than the increase in the network capacity, but the current solutions fail to guarantee even the promised achievable transfer throughputs. In this paper, we propose a novel dynamic throughput optimization model based on mathematical modeling with offline knowledge discovery/analysis and adaptive online decision making. In offline analysis, we mine historical transfer logs to perform knowledge discovery about the transfer characteristics. Online phase uses the discovered knowledge from the offline analysis along with real-time investigation of the network condition to optimize the protocol parameters. As real-time investigation is expensive and provides partial knowledge about the current network status, our model uses historical knowledge about the network and data to reduce the real-time investigation overhead while ensuring near optimal throughput for each transfer. Our novel approach is tested over different networks with different datasets and outperformed its closest competitor by 1.7x and the default case by 5x. It also achieved up to 93\% accuracy compared with the optimal achievable throughput possible on those networks.
\end{abstract}

}

\maketitle

\IEEEdisplaynontitleabstractindextext

%
\IEEEpeerreviewmaketitle

\IEEEraisesectionheading{\section{Introduction}
\label{sec:introduction}}
\IEEEPARstart{A}{pplications} in a variety of spaces --- scientific, industrial, and personal --- now generate
more data than ever before.
Large scientific experiments, such as 
high-energy physics simulations~\cite{CMS, ATLAS},
climate modeling~\cite{Climate, easterling2000climate},
environmental and coastal hazard prediction~\cite{Klein200335, carrara1999use},  
genomics~\cite{BLAST, morozova2008applications}, 
and astronomic surveys~\cite{loredo2007analyzing, eisenstein2011sdss}
generate data volumes reaching several Petabytes per year. Data collected from remote sensors and satellites, dynamic data-driven applications, digital libraries and preservations are also producing extremely large datasets for real-time or offline processing~\cite{Ceyhan07, Tummala07}. 
With the emergence of social media, video over IP, and more recently the trend for Internet of Things (IoT), we see a similar trend in the commercial applications as well, and 
it is estimated that, in 2017, more IP traffic will traverse global networks than all prior ``Internet years" combined. The global IP traffic is expected to reach an annual rate of 1.4 zettabytes, which corresponds to 
 nearly 1 billion DVDs of data transfer per day for the entire year~\cite{Cisco_2016}.
 
As data becomes more abundant and data resources become more heterogenous, 
accessing, sharing and disseminating these data sets become a bigger challenge.
Managed file transfer (MFT) services such as Globus~\cite{globusonline}, PhEDEx~\cite{egeland2010phedex}, Mover.IO~\cite{moverio}, and B2SHARE~\cite{ardestani2015b2share} have allowed users to easily move their data, but these services still rely on the users providing specific details to control this process, and they suffer from inefficient utilization of the available network bandwidth and far-from-optimal end-to-end data transfer rates.
%
%
%
End-to-end data transfer performance can be significantly improved by tuning the application-layer transfer protocol parameters (such as pipelining, parallelism, and concurrency levels). Sub-optimal choice of these parameters can lead to under-utilization of the network or may introduce link congestion, queuing delays, packet loss, and  end-system over-utilization. It is hard for the end users to decide on optimal levels of these parameters statically, since  static setting of these parameters might prove sub-optimal due to the dynamic nature of the network which is very common in a shared environment. 

In this paper, we propose a novel two-phase dynamic transfer throughput optimization model for big data based on mathematical modeling with offline knowledge discovery/analysis and adaptive online decision making. During the  offline analysis phase, we mine historical transfer logs to perform knowledge discovery about the transfer characteristics. During the online phase, we use the discovered knowledge from the offline analysis along with real-time investigation of the network condition to optimize the protocol parameters. As real-time investigation is expensive and provides partial knowledge about the current network status, our model uses historical knowledge about the network and data to reduce the real-time investigation overhead while ensuring near optimal throughput for each transfer. We have tested our network and data agnostic solution over different networks and observed up to 93\% accuracy compared with the optimal achievable throughput possible on those networks. Extensive experimentation and comparison with best known existing solutions in this area revealed that our model outperforms existing solutions in terms of accuracy, convergence speed, and achieved end-to-end data transfer throughput. 

 In summary, the contributions of this paper include:
 (1) it performs end-to-end big data transfer optimization completely at the application-layer, without any need to chance the existing infrastructure nor to the low-level networking stack;
 (2) it combines offline knowledge discovery with adaptive real-time sampling to achieve close-to-optimal end-to-end data transfer throughput with very low sampling overhead;
  (3) it constructs all possible throughput surfaces in the historical transfer logs using cubic spline interpolation, and creates a  probabilistic confidence region with Gaussian distribution to encompass each surface;
(4) in real time, it applies adaptive sampling over the pre-computed throughput surfaces to provide faster convergence towards maximally achievable throughput;
(5) it outperforms state-of-the-art solutions in this area in terms of accuracy, convergence speed, and achieved throughput.

The rest of the paper is organized as follows: 
Section II gives background information; 
Section III presents the problem formulation; Section IV discusses our proposed model;
Section V presents the evaluation of our model; Section
VI describes the related work in this field; and Section VII
concludes the paper with a discussion on the future work.

\section{Background}
\label{sec:background}
Wide-area data transfers perform poorly in a high-speed long RTT network with the existing protocol stack. Updating the current protocols are expensive, and the adaptation of new protocol is very slow. Moreover, any protocol update requires kernel-space modifications. Operating systems developers take a long time to introduce such kernel updates, and sometimes reluctant to such problem-specific kernel updates. Application layer solutions are free from problems as mentioned earlier. These solutions can be deployed in the user-space instead of kernel-space update. Application layer can access tunable protocol parameters from the underlying protocol stack. Existing protocol stack mostly lacks those parameters and their tuning. The protocol parameters (i.e., concurrency, parallelism, pipelining, buffer size) have a significant impact on data transfer efficiency.  
A brief description of these parameters are given below.  


\textbf{Concurrency} ($cc$) refers to the task level parallelism. It controls the number of server processes where each process can transfer individual file. It can accelerate the transfer throughput when a large number of files needs to be transferred. Concurrency can also take advantage of Parallel file systems (e.g. GPFS, Lustre) with multiple concurrent servers with meta-data management. 

Each server process can transfer a different portion of a file in parallel. We define the number of parallel streams for each server as \textbf{Parallelism} ( $p$). It is a good choice to transfer medium and large files. We can get the full performance of Parallelism with Parallel File Systems where files are divided and distributed on different disks. Therefore, the number of total parallel data streams is $(cc\times p)$. Increasing number of parallel data streams can increase the achievable throughput, however, excessive use of streams might lead to packet loss and force TCP to decrease the sending rate with congestion avoidance algorithm. 

Control channel idleness is a major bottleneck to transfer large number of  files. After each file transfer, server sends an acknowledgement to initiate the next file transfer. This acknowledgement can take up to 1 Round-Trip-Time (RTT) between each transfer and hurt the overall throughput significantly in a long RTT network. Moreover, TCP will shrink window size to zero if it detects data channel idleness. These issues can be solved by queuing multiple file transfer requests without waiting for the acknowledgements. This technique can transfer large block files like a single large file. We define the size of the outstanding file transfer request queue as \textbf{Pipelining} ( $pp$).

In a shared network infrastructure, other contending transfers also compete for network resources to achieve their performance goal.
Even though the increased number of data streams ($cc \times p$) opened for a transfer job can improve the performance significantly, an excessive number of streams can introduce congestion. 
The congestion becomes direr when other contending transfer jobs unilaterally increase the number of streams to get an increased share of the network.  
Therefore, we need to analyze the impact of parameters on data transfer performance. 

\subsubsection{Analysis of parameters}

\begin{figure*}[t]
    \centering
    \begin{subfigure}[t]{0.3\textwidth}
    \centering
        \includegraphics[height=45mm,width=55mm]{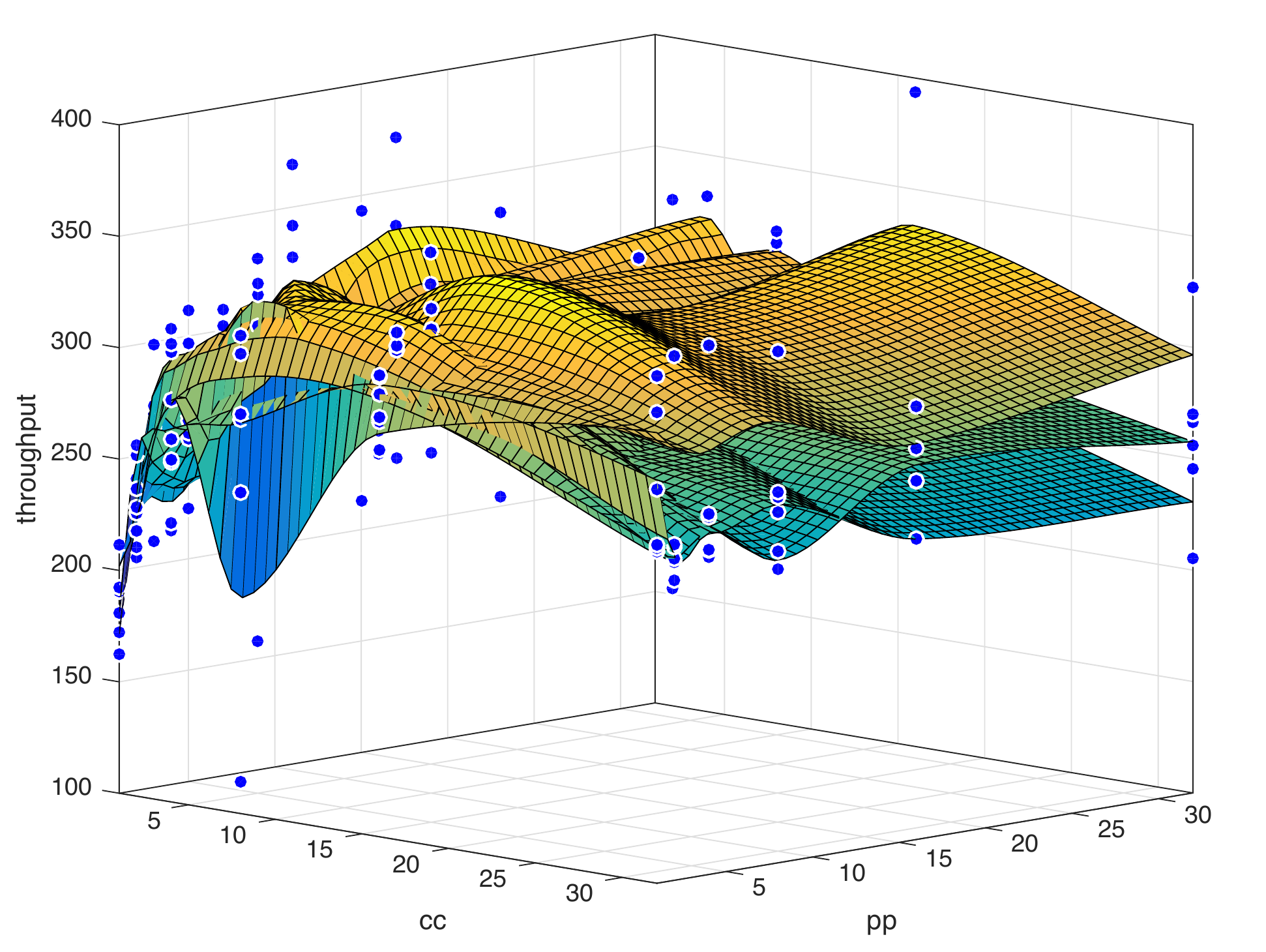}
        \caption{Small-sized Dataset}
    \end{subfigure}
    ~
    \begin{subfigure}[t]{0.3\textwidth}
    	\centering
        \includegraphics[height=45mm,width=55mm]{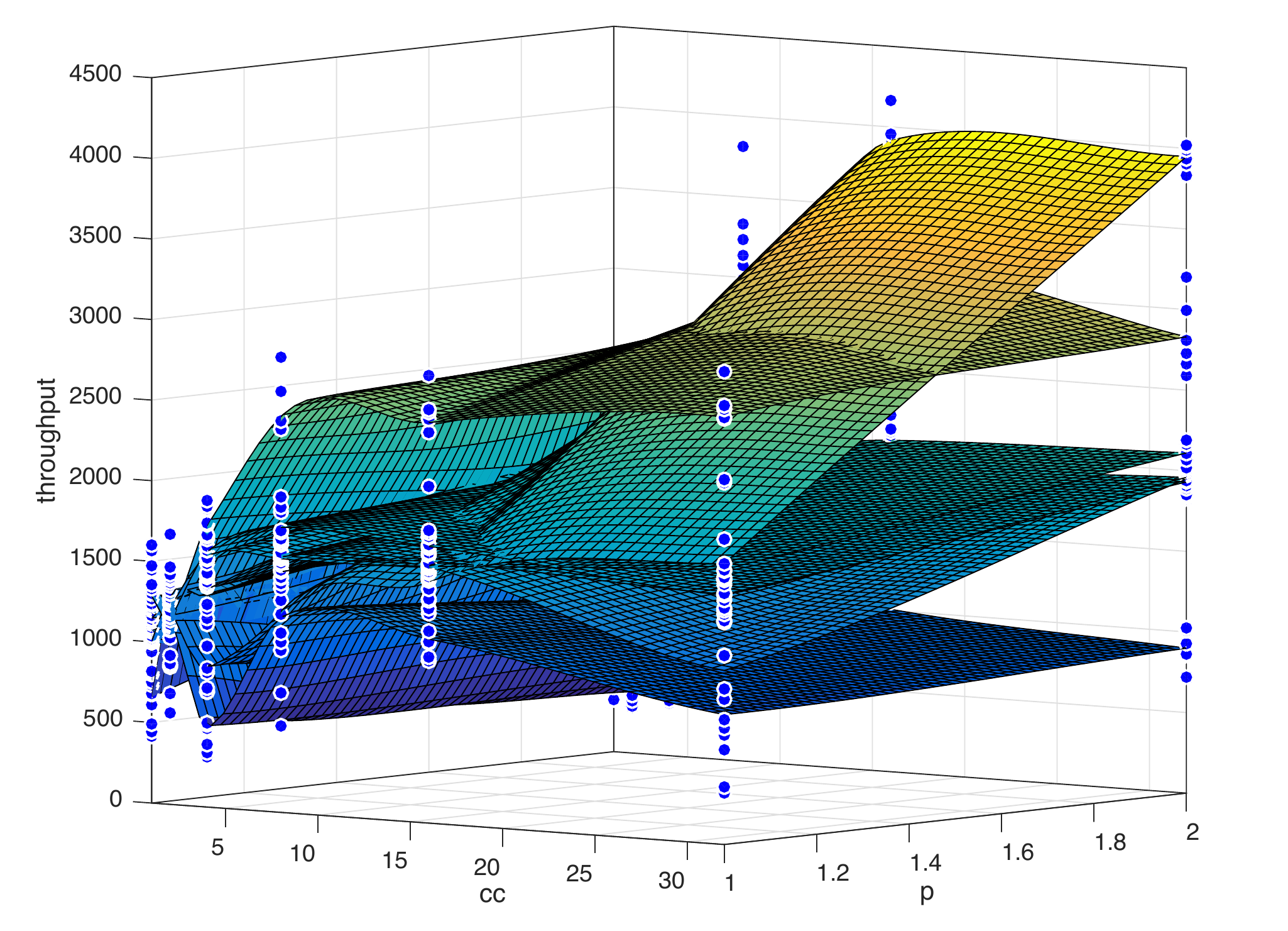}
        \caption{Medium-sized Dataset}
    \end{subfigure}
    ~ 
    \begin{subfigure}[t]{0.3\textwidth}
    \centering
        \includegraphics[height=45mm,width=55mm]{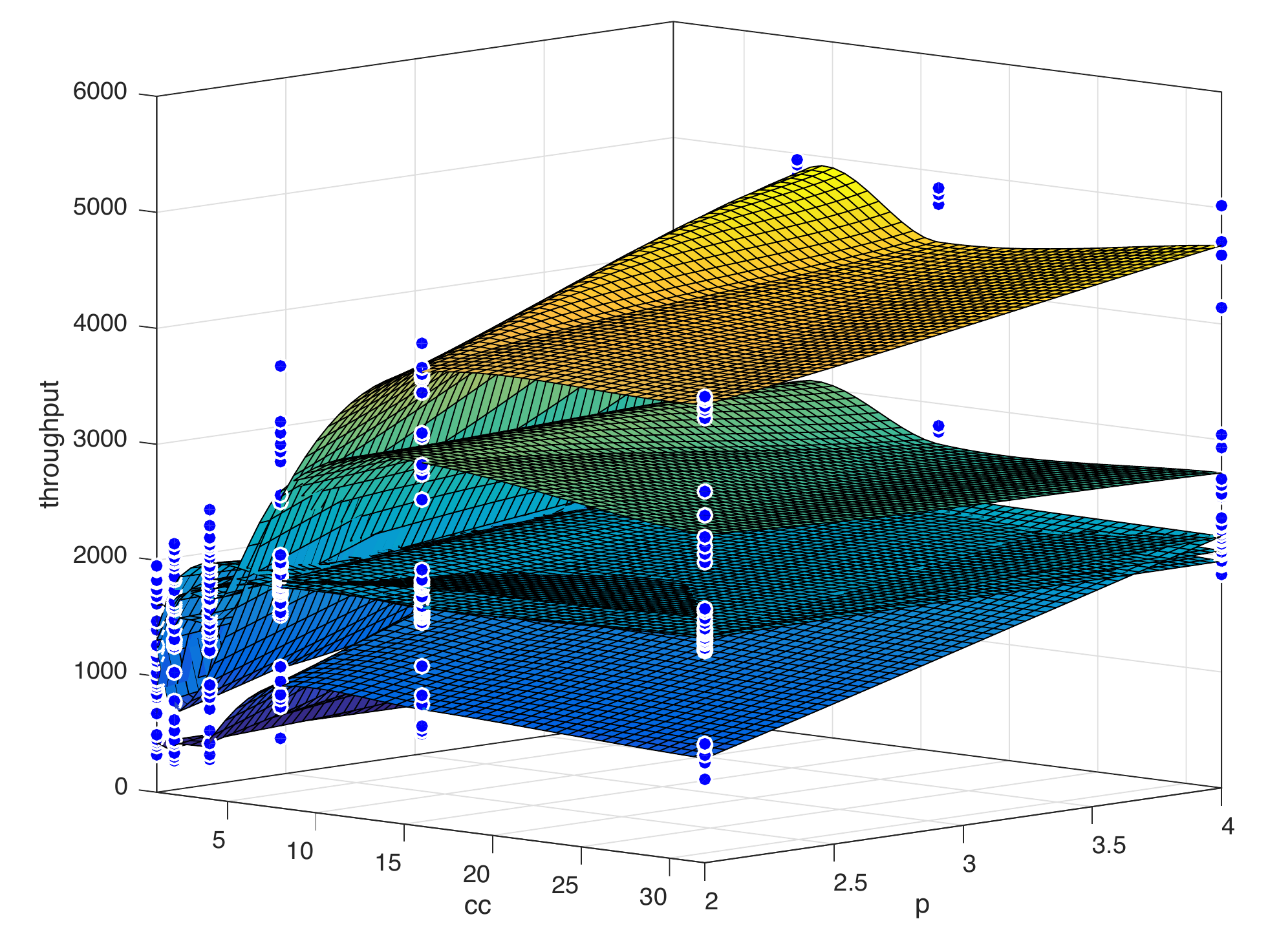}
        \caption{Large-sized Dataset}
    \end{subfigure}
     \caption{Piece-wise cubic interpolation surface construction.}
     \label{fig:3diinterpolation}
 \end{figure*}


Large scale wide area transfers can take longer time to finish (hours or days). 
During the transfer time network condition might change drastically. 
Link capacity reduction due to maintenance or failure could change network bandwidth. Moreover, the available bandwidth also changes as the other contending transfers come and go. 
Therefore, we need a mechanism to dynamically tune the protocol parameters to adapt with the changing network conditions. 
We also need to ensure that users can receive a fair share of the network bandwidth. 
\begin{figure}[t]
\begin{centering}
\includegraphics[keepaspectratio=true,angle=0,width=95mm]{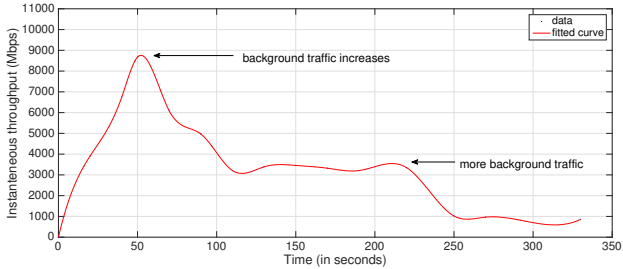}
	\end{centering}
    \vspace{-3mm}
\caption{ Achievable throughput (Gbps) for different models in a multi-user scenario in Chameleon Cloud between a CHI-UC and a TACC node.} \label{fig:multiuser-case}
\vspace{-5mm}
\end{figure}


\section{Problem Formulation}
\label{sec:problemformulation}
Optimal choice of application level parameters are necessary to achieve high data transfer throughput in a long RTT WAN. 
As we have seen in Section, achievable transfer throughput depends on application level parameters, network characteristics, and the data itself. 
However, in a shared network, the data transfer job has to compete with other contending transfers. 
The performance of the data transfer job varies with the transfer load of the contending transfers.  
%
Figure shows the performance of a data transfer job with different background traffic. 
We can conclude that optimal parameter for a background traffic can overburden the network when the background traffic load increases. 
Similarly, the aforementioned optimal parameter can severely under-utilize the network when background traffic load decreases.
Therefore, we need to consider the external traffic load along with aforementioned factors to model the achievable throughput.

Given a source endpoint $e_s$ and destination endpoint $e_d$ with a link bandwidth $b$ and round trip time $rtt$; a dataset with a total size $f_{all}$, average file size $f_{avg}$,and  number of files $n$; and set of protocol parameters $\theta=\{cc, p, pp\}$, the throughput $th$ optimization problem can be defined as:

\begin{equation}
th = f(e_s,e_d,b,rtt,f_{avg}, n,cc,p,pp,l_{ctd})
\end{equation}
We define the load from contending transfers as $l_{ctd}$. 

Our goal is to maximize the data transfer throughput using parameter values that are optimal for the network condition and dataset. We define our optimization problem as follows.

\begin{equation}
\begin{array}{ll}
\underset{ \{cc,p,pp\} }{\mathrm{argmax}} &  \displaystyle\int_{t_s}^{t_f} th \\
\text{subject to.} & cc \times p \leq \mathbb{N}_{streams} \\ 
				   & pp \leq \mathbb{P} \\
                   & th \leq b \\
\end{array}
\end{equation}

where $t_s$ and $t_f$ are the transfer start and end times respectively. Additionally, $\mathbb{N}_{streams}$ and $\mathbb{P}$ are the maximum allowable parameter values in the network. 

In a distributed and shared environment, each user need to maximize their throughput without hurting the performance of other contending transfers. Therefore, we need another constraint to ensure fairness among the contending transfers. 

Sometimes large scale transfers take hours to complete. Network conditions and external traffic load might change during the transfer. Assuming that the data transfer started with optimal parameter values might become sub-optimal during the transfer. Therefore, we need a mechanism to detect the network condition change during the transfer and adapt the parameters in real-time to achieve optimal performance in changed condition.

Data transfer optimization problem can be addressed in two ways - (1) Centralized approach, (2) Distributed approach. In a centralized approach, a central scheduler can distribute the parameters to contending transfers. This approach is applicable when both source and destination and the link connecting them are managed by single administrative domain. It has a global view of the network and contending transfer. Therefore, scheduling decisions are precise. Centralized approach can be used as a service where compute resources and their network connections are managed by a single administrative domain. We also introduce a distributed approach that can be used by standalone users where the data path is controlled by different administrative domains. In this approach user can sense the network and converge to receive a fair share of the network. However, lack of the global network view can induce oscillation to choose parameters and could achieve slightly less throughput compare to centralized approach.  

\section{Model Overview}
\label{sec:modeloverview}
%
In a shared environment, each user tries to maximize their throughput without hurting other contending transfers. 
End users are unaware of the characteristics of the other contending transfers. 
Therefore, each user needs to check the available bandwidth to set the parameters. 
Moreover, the user needs a mechanism to tune the parameters and ensure fairness constraint dynamically. 
At first, we describe the simplest solution for the distributed approach solve the problem and explain its inefficiencies. 
In this solution:

\begin{itemize}
	\item Each user starts the transfer with default parameter values. 
    \item Then periodically, the user increases the protocol parameters with constant steps until it detects queuing delay.
    \item Increasing queuing delay indicates upcoming congestion. 
    \item Therefore, the user can cut back the parameter values up to a certain level.
    \item In case of packet loss (indicate congestion), the user reduces parameter values more drastically. 
\end{itemize}

We identify three major performance issues in this simple solution. 

\textbf{Issue 1:} Protocol parameters are heavily dependent on end systems, network links, type of data, and application that performs data transfer. 
Different transfers need different protocol parameters to achieve optimal performance. 
The default parameter values could become sub-optimal at the beginning of the transfer. 
Therefore, we need an approach that can provide protocol parameters to individual transfers in a more personalized way.

\textbf{Issue 2:} The periodic additive increase of protocol parameters during the transfer could introduce extra overhead. 
When initial parameter choices are far away from optimal parameters, additive increase could be slower and  the transfer takes a longer time to reach the optimal parameter level. 
Moreover, the increase in concurrency/parallelism opens new TCP streams that go through TCP slow start phase. 
Therefore, we need an approach where protocol parameters can converge towards optimal level faster. 

\textbf{Issue 3:} During congestion, the user reduces the parameters up to a predefined level. Too aggressive reduction of parameters can reduce performance. 
When congestion is over, transfer needs a longer time to reach the optimal level again. 
Therefore, we need a mechanism to reduce parameters up to a level that is good enough to clear congestion and close enough to optimal level. 
This strategy helps to regain optimal performance when congestion resolved. 

\textbf{Issue 4:} This simplistic approach does not provide any mechanism for fairness. 


To address these issues, we proposed a hybrid model that perform historical analysis on data transfer logs along with real-time tuning of the parameters during the transfer. 
Our historical log analysis based approach to provide more personalized 
parameter native to the end user system and network. 
This approach can resolve issue (1) explained earlier. 
This historical log analysis provides parameter settings that are optimal for the system and the network condition. 
It also reduces the slow convergence problem explained in Issue -2. 
In Section V, we show that the convergence speed of our approach is much better than default or heuristics based approaches. 
However, historical log analysis induces extra overhead of processing, while the default parameter settings or heuristic based approaches does not have such overhead. 
Therefore, we decided to perform historical analysis during offline. The overhead of offline optimization can be amortized over many subsequent transfers that can benefit from such parameter analysis. The optimal results are stored as key-value pair. 

The offline analysis module is an additive model. That means when new logs are generated for a certain period of time, we do not need to combine them with previous logs and perform analysis on the entire log \textit{(old log + new log)} from scratch. Data transfer logs can be collected for a certain period of time and then the additive offline analysis can be performed on those new logs only. For services like Globus, historical logs can be analyzed by a dedicated server and results can be shared by the users. 

Initially, the main transfer process perform a sample transfer to detect the network conditions, such as RTT, Queuing delay, packet loss rate, etc. The sample transfer is performed using a small predefined portion of the data that needs to be transfered. Then it gets parameter settings for such dataset and network conditions from the offline optimization module. The results are already precomputed in offline module, therefore, can be retrieved in constant time. Main transfer process starts transferring the rest of the data using these parameter settings. Then it periodically monitors the health of the data transfer. Whenever, it detects persistent change in network condition and external traffic load, it asks offline optimization module for new parameters that are optimal to the current condition.

\subsection{Offline Optimization}
\label{subsec:Offlineoptimization}

\begin{figure*}[t]
	\centering
    \includegraphics[keepaspectratio=true,width=150mm]{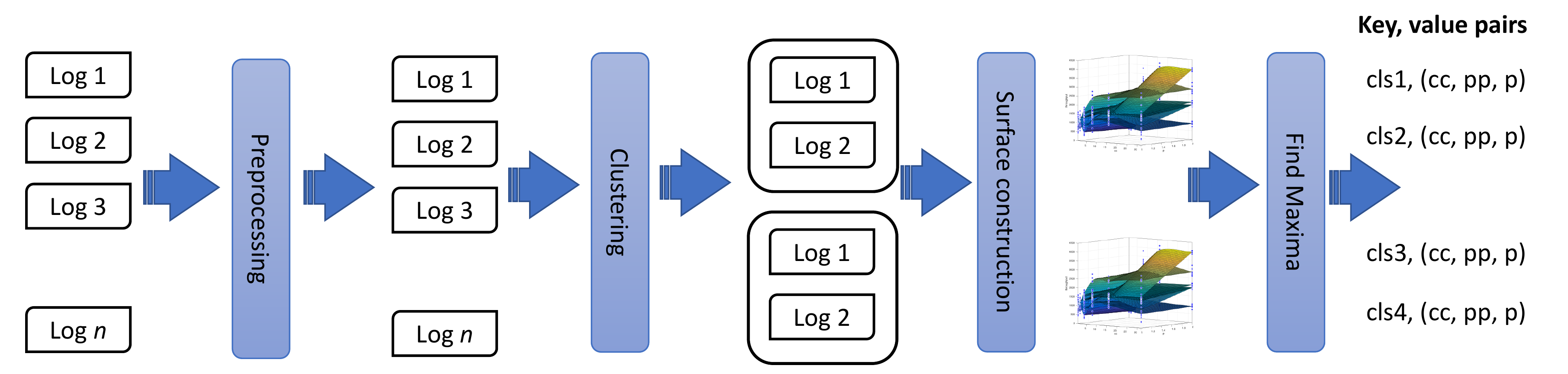}
    \caption{Flow of historical log through offline optimization}
    \label{fig:3diinterpolation}
\end{figure*}

Offline analysis collects useful information from the historical logs so that those information can be used by the online module to converge faster. Offline analysis consists of five phases: \textit{(i)} clustering logs in hierarchy; \textit{(ii)} surface construction; \textit{(iii)} finding maximal parameter setting; \textit{(iv)} accounting for known contending transfers; and \textit{(v)}identifying suitable sampling regions. 

\subsubsection{Clustering logs}
\label{subsub:clusteringlogs}
Historical data transfer logs contain information about the verity of transfers performed by the users. Therefore, a natural approach would be cluster the logs based on different matrices. Assuming that we have a historical log, $L$ of $n_{log}$ log entries, We can define our clustering problem as $(L,m)$, where $m$ is the number of target clusters. The clusters of the historical logs are $C = \{ C_{1},...,C_{m}\}$, where $\{n_{1},...,n_{m}\}$ denote the sizes of the corresponding clusters. We consider a pairwise distance function, $\it{d}(x,x^\prime)$ where $x,x^\prime \in L$. We have tested clustering algorithm for different pair-wise distance functions. For clustering, we have tested two well-known approaches: (1) K-means++~\cite{kmeans++}; (2) Hierarchical Agglomerative  Clustering (HAC) with Unweighted Pair Group Method with Arithmetic Mean (UPGMA)~\cite{upgma_clustering}. K-means clustering algorithm suffers from initial centroid selection, and wrong initialization could lead to wrong clustering decision. However, K-means++ provides a theoretical guarantee to find a solution that is $O(log\ m)$ competitive to the optimal K-means solution. For HAC, distance between two clusters $C_{i}$ and $C_{j}$ is defined as $D(C_{i},C_{j})$ in Equation (\ref{eq:UPGMAdistance}):

\begin{equation}
\label{eq:UPGMAdistance}
D(C_{i},C_{j}) = d(c_i,c_j) = \sqrt[]{(c_i - c_j)^{2}}
\end{equation}

Where $c_i$ and $c_j$ are the corresponding cluster centroid of $C_i$ and $C_j$. HAC computes proximity matrix for the initial clusters and combines two clusters with minimum distance $D(C_i,C_j)$. Then, it updates the rows and columns of proximity matrix with new clusters and fills out the matrix with new $D(C_i,C_j)$. This process is repeated until all clusters are merged into a single cluster. 

Clustering accuracy depends on the appropriate number of clusters $k$. In this work, we have used Calinski and Harabasz index (CH index) to identify the appropriate number of clusters. CH index can be computed as:
\begin{equation}
CH(m) = \dfrac{\Phi_{inter}(m)/(m-1)}{\Phi_{inter}(m)/(n-m)}
\end{equation}
Where $\Phi_{inter}$ is the ``between-cluster" variation and $\Phi_{intra}$ is the ``within-cluster"" variation. Both can be defined as the sum of Euclidean distance as bellow:

\begin{equation}
 \Phi_{inter}(m) = \sum^{m}_{i=1} \sum_{x \in c_i} (x - c_i)^2
\end{equation}

\begin{equation}
 \Phi_{intra}(m) = \sum^{m}_{i=1} n_k(\bar{C_k} - \bar{x})^2
\end{equation}
Where $\bar{C_k}$ is the mean of points in cluster $k$ and $\bar{x}$ is the overall mean. Largest $CH(m)$ score is preferable.

\subsubsection{Surface Construction}
\label{subsubsec:surfaceconstruction}
Achievable throughput for a given cluster $C_{i}$ can be modeled as a polynomial surface which depends on the protocol tuning parameters. We have tried three models to see how accurate those can capture the throughput behavior. The models are: (1) quadratic regression; (2) cubic regression; and (3) piecewise cubic interpolation. 

\textbf{\textit{ Piecewise cubic spline interpolation.}} We model throughput with cubic spline surface interpolation \cite{kincaid2009numerical}. Before introducing the interpolation method, we should explain the relationship among the parameters briefly. Concurrency and pipelining are responsible for a total number of data streams during the transfer, whereas, pipelining is responsible for removing the delay imposed by small files. Due to their difference in characteristic, we model them separately. At first, we construct a $2$-dimension cubic spline interpolation for $g(pp) = th$. Given a group of discrete points in $2$-dimension space $\{(pp_i,th_i)\}, i = 0,...,N$, the cubic spline interpolation is to construct the interpolant $g(pp) = th$ by using piecewise cubic polynomial $g_i(pp)$ to connect between the consecutive pair of points $(pp_i,th_i)$ and $(pp_{i+1},th_{i+1})$. The coefficients of cubic polynomials are constrained to guarantee the smoothness of the reconstructed curve. This is implemented by controlling the second derivatives since each piecewise relaxed cubic polynomial $g_i$ has zero second derivative at the endpoints. Now we can define each cubic polynomial piece as:  
\begin{equation}
g_i(pp) = c_{i,0} + c_{i,1}pp + c_{i,2}pp^2 + c_{i,3}pp^3,\forall pp \in [pp_i,pp_{i+1}].
\end{equation}

Periodic boundaries can be assumed as $g(pp_{i+1}) = g(pp_i)$. Coefficients $c_{i,j}$, where $j = 1,2,3$, of piece-wise polynomial $g_i(pp)$ contains $4(N-1)$ unknowns. We can have:
\begin{equation}
g_i(pp_i) = th_i, \quad i = 1,...,N
\end{equation}
Hence, the $N$ continuity constraints of $g(pp)$ are as:
\begin{equation} \label{eq:constraint2}
g_{i-1}(pp_i) = th_i = g_{i}(pp_i), \quad i = 2,...,N.
\end{equation}
We can get $(N-2)$ constraints from Equation (\ref{eq:constraint2}) as well. We can impose additional continuity constraints up to second derivatives. 
\begin{equation} \label{eq:constraint3}
\dfrac{d^2 g_{i-1}}{d^2 pp} (pp_i) = \dfrac{d^2 g_{i}}{d^2 pp} (pp_i), \quad i = 2,...,N
\end{equation}
We can get $2(N-2)$ constraints from Equation (\ref{eq:constraint3}). The boundary condition for relaxed spline could be written as:

\begin{equation} \label{eq:constraint4}
\dfrac{d^2 g}{d^2 pp} (pp_1) =  \dfrac{d^2 g}{d^2 pp} (pp_n) = 0
\end{equation}

So we have $N + (N-2) + 2(N-2) +2 = 4(N-1)$ constraints in hand. The coefficients can be computed by solving the system of linear equations. This example is extended to generate surface with two independent variables.

Then, we model throughput as a piece-wise cubic spline surface. It can be done by extending the $2$-dimension cubic interpolation scheme presented above  to model the function of throughput. A short overview is given below:
We first fix the value of $pp$.
The throughput $f(p,pp,cc)$ then becomes $f_{pp}(p,cc)$
which is a surface in $\Psi^2 \times \mathbb{R}$. Where $\Psi$ domain of parameters $\{cc, p, pp\}$.  
Data points can be represented as
$P = \{ (p_{i},cc_{j},th_{i,j}) | i = 1,2,\ldots,N, j = 1,2,\dots,M, (x_i,y_j) \in G \}$,
where $G$ is an $N \times M$ rectangle grid,
and each $(p_{i},cc_{j})$ denotes the grid point at $i$-th row and $j$-th column. The piecewise cubic interpolation method will construct
an interpolated cubic function $f_{r(i,j)}(p,cc)$
for each rectangle $r(i,j)$ from the grid $G$,
where $r(i,j)$ rectangle can be defined by $[p_i, p_{i+1}] \times [cc_j, cc_{j+1}]$.
$f_{r(i,j)}(p,cc)$ should fit the $th$-values of $P$ at the vertices of $r(i,j)$, \emph{i.e.}

\begin{gather*} 
f_{r(i,j)}(p_i,cc_j) = th_{i,j} \\
f_{r(i,j)}(p_{i+1},cc_j) = th_{i+1,j} \\
f_{r(i,j)}(p_i,cc_{j+1}) = th_{i,j+1} \\
f_{r(i,j)}(p_{i+1},cc_{j+1}) = th_{i+1,j+1} \\
\end{gather*} 

The interpolated cubic functions should also maintain smoothness at grid points. The continuity constraints are computed as an extension of   Equation (\ref{eq:constraint3}) and  (\ref{eq:constraint4}). 

Formally, this means that there exist numbers
$\Omega_{1}(i,j), \Omega_{2}(i,j), \Omega_{1,2}(i,j)$ for each $i = 1,2,\ldots,N, j = 1,2,\dots,M$,
such that for each $rect(i,j)$, and each vertex $(p_{i'},cc_{j'})$ of $rect(i,j)$,
\[
(\partial f_{r(i,j)}(p_i,cc_j) / \partial p)|_{p_{i'},cc_{j'}} = D_{1}(i',j')
\]
\[
(\partial f_{r(i,j)}(p_i,cc_j) / \partial cc)|_{p_{i'},cc_{j'}} = D_{2}(i',j')
\]
\[
(\partial f_{r(i,j)}(p_i,cc_j) / \partial p \partial cc)|_{p_{i'},cc_{j'}} = D_{1,2}(i',j')
\]

\begin{figure}[t]
    \centering
    \begin{subfigure}[t]{0.2\textwidth}
    	\centering
        \hspace{-1.2cm}
        \includegraphics[keepaspectratio=true,width=45mm]{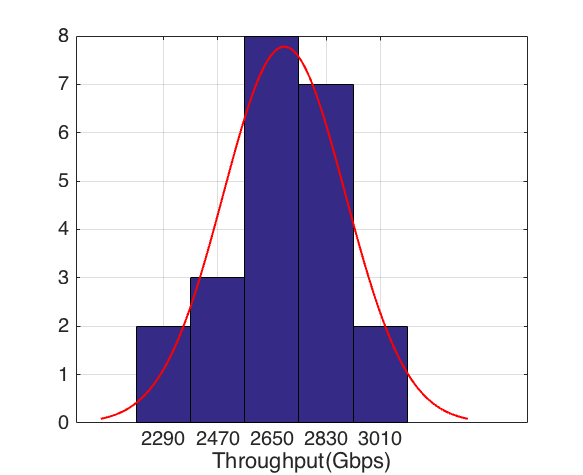}
        \caption{}
    \end{subfigure}%
    ~ 
    \begin{subfigure}[t]{0.2\textwidth}
    \centering
        \includegraphics[keepaspectratio=true,width=44mm]{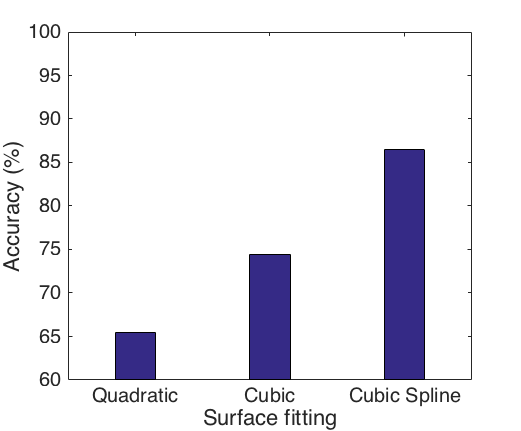}
        \caption{}
    \end{subfigure}
    \caption{(a) Distribution of throughput values under similar external loads; b) Accuracy of different surface construction methods.}
    \label{fig:surface_accuracy}
\end{figure}

After solving the  constraints (which is a linear system),
all the functions $f_{rect(i,j)}(x,y)$ where $i = 1,2,\ldots,N-1, j = 1,2,\dots,M-1$
form a piecewise smooth function which fixes data point set $P$.
Figure \ref{fig:3diinterpolation} shows the constructed piece-wise cubic surfaces of throughput for different $cc$ and $p$ values. We can see from the graph surfaces for small files are more complex than the medium and large file. 
Figure \ref{fig:surface_accuracy}(b) shows the accuracy of different surface construction models. It can be seen that piece-wise cubic spline outperforms all other models and achieves almost 85\% accuracy.

Data transfer requests within the same cluster $C_{i}$ with the same protocol parameter values might have a deviation from one another due to measurement errors and many other network uncertainties such as different packet route in the network layer and minor queuing delay. 
We define those data points with the same protocol parameter entries as $\omega$. To model this deviation, we have used a Gaussian confidence region around each constructed surface. The probability density function of a Gaussian distribution is: 

\begin{eqnarray}
&& p(\omega;\mu,\sigma) = \dfrac{1}{\sqrt{2\pi\sigma^2}}e^{-\dfrac{(\omega-\mu)^2}{2\sigma^2}}, \label{eq:confidence}\\
&& \mu = \frac{1}{N}\sum_{i=1}^N th_i, \label{eq:confidence:mean}\\
&& \delta = \sqrt{\frac{1}{N}\sum_{i=1}^N(th_i-\mu)^2}, \label{eq:confidence:var}
\end{eqnarray}
where $\mu$ is the mean and $\sigma$ is the standard deviation of the data distribution. Figure \ref{fig:surface_accuracy}(a) shows the data model for the Gaussian distribution.

\subsubsection{Finding Maximal Parameters}
\label{subsec:findingmaximalparams}
Very high protocol parameter values might overburden the system. For this reason, many systems set upper bound on those parameters. Therefore, the parameter search space has a bounded integer domain. Assuming $\beta$ is the upper bound of the parameters, cubic spline surface functions can be expressed as $f_i: \Psi^{3} \Rightarrow \mathbb{R}^{+}$, where $\Psi = \{1,2,..,\beta\}$. To find the surface maxima, we need to generate all local maxima of $F = \{f_1,...,f_p\}$. This is achieved by performing the second partial derivative test on each $f_k$ \cite{kincaid2009numerical}. The main idea is presented below. 

First, we calculate the Hessian matrix of $f_k$ as:
\begin{equation}
\label{eq:Hessian_fk}
H_k = J(\bigtriangledown f_k) =\left[ {\begin{array}{*{20}c}
   \dfrac{\partial^2 f_k}{\partial p^2} & \dfrac{\partial^2 f_k}{\partial p\ \partial cc} & \dfrac{\partial^2 f_k}{\partial p\ \partial pp}  \\
   ... & ... & ... \\
\dfrac{\partial^2 f_k}{\partial pp\ \partial p} & \dfrac{\partial^2 f_k}{\partial pp\ \partial cc} & \dfrac{\partial^2 f_k}{\partial pp^2}  \\  
 \end{array} } \right],
\end{equation}
where $J$ stands for the Jacobian matrix:
\begin{equation}
J(F) = \left[ {\begin{array}{*{20}c}
   \dfrac{\partial f_1}{\partial p} & \dfrac{\partial f_1}{\partial cc} & \dfrac{\partial f_1}{\partial pp}  \\
   ... & ... & ... \\
\dfrac{\partial f_p}{\partial p} & \dfrac{\partial f_p}{\partial cc} & \dfrac{\partial f_p}{\partial pp}  \\  
 \end{array} } \right]
\end{equation}

Then we obtain the coordinates of all local maxima in $f_k$ by calculating the corresponding $\{p,pp,cc\}$'s such that $H_k(p,pp,cc)$ is negative definite. Hence, the set of local maxima of $f_k$ is obtained. Finally, the surface maxima is generated by taking the maximum among all local maxima sets of $F$. 



\subsubsection{Identifying Suitable Sampling Regions}
\label{subsec:unknownload}
Identifying the suitable sampling region is a crucial phase that helps online adaptive sampling module to converge faster. 
However, not all the regions on a surface are interesting. Many parameter coordinates of a surface are suboptimal. We are interested in regions which have a better possibility of achieving high throughput. The regions containing distinguishable characteristics of the surfaces and containing the local maxima of those surfaces are more compelling. Exploring those regions could lead to a near-optimal solution much faster. Assume the cluster $C_{i}$ contains $\eta$ number of the surfaces that can be written as $S = {f_{1}, ... f_{\eta}}$. Now, we can extract the neighborhood with a predefined radius $r_d$ that contains maxima for all the surfaces in $S$. Assume the set $R_m$ contains all those neighborhood of maxima. We are also interested in regions where surfaces are clearly distinguishable. The goal is to find the regions where surfaces are maximally distant from one another. This problem can be formulated as a max-min problem. Selection can be done by taking the maximum of all pair shortest distance between the surfaces. To achieve that we perform uniform sampling $u = \{u_{1},...,u_{\gamma}\}$ from surface coordinate $(p,cc,pp)$  for surfaces in $S$. Therefore, $u$ could be written as: 

\begin{equation}
u = \{u_{1},...,u_{\gamma}\} =  \{(p_i,cc_i,pp_i)\}^{\gamma}_{i=1} 
\end{equation}
We define $\Delta^{min}_{u_i}$ as the minimum distance between any two pair of surfaces that can be expressed as: 
\begin{equation}
\label{eq:region_selection}
\forall u_{k} \in u, \quad \Delta^{min}_{u_k} = \underset{ \forall i,j \in \{1,...,\eta\} }{\mathrm{min}} |f_{i}(u_k) - f_{j}(u_{k})| \quad \textrm{where} \quad i \neq j
\end{equation}
 
 After sorting the list in descending order we choose, $\lambda$ $(1<\lambda < k)$, number of the initial samples from the sorted list.  
 Assume the set of points we get after solving the Equation (\ref{eq:region_selection}) is $R_c$. We define suitable sampling region as :
 \begin{equation}
R_s = R_m \cup R_c
\end{equation}
During online analysis, we will use the region in $R_s$ to perform the sample transfers.


\subsection{Dynamic Control}
\label{sub:dynamiccontrol}
This module is initiated when a user starts a data transfer request. Adaptive sampling is dependent on online measurements of network characteristics. It is essential to assess the dynamic nature of the network that is helpful to find the optimal parameter settings. A sample transfer could be performed to see how much throughput it can achieve. However, a single sample transfer could be error prone and might not provide clear direction towards the optimal solution. Our algorithm adapts as it performs sample transfers by taking guidance from offline surface information. This approach can provide faster convergence. An overview of the module is presented in Algorithm (\ref{algo:onlinesampling}).  Online module queries the offline analysis module with network and dataset characteristics. Offline module finds the closest cluster and returns the throughput surfaces along with associated external load intensity information and suitable sampling region for each surface. 



The online module sorts the surfaces in descending order based on external load intensity value (Lines 17-18).  
Adaptive sampling module takes the dataset that is needed to be transferred and starts performing sample transfers from the dataset. To perform the first sample transfer, the algorithm chooses the surface with median load intensity, $f_{median}$, and performs the transfer with: 

\begin{equation}
\theta_{s,median} = \{p,cc,pp\} = \mathrm{argmax}(f_{s,median})
\end{equation}

\IncMargin{1em}
\begin{algorithm}[t]
	\SetKw{in}{in}
	\SetKwData{Left}{left}\SetKwData{This}{this}
    \SetKwData{Up}{up}\SetKwData{Log}{Log}
    \SetKwData{ismedian}{$e_{s,median}$}\SetKwFunction{Median}{Median}
    \SetKwFunction{Union}{Union}
    \SetKwFunction{FindCompress}{FindCompress}
    \SetKwFunction{FindClosestSurface}{FindClosestSurface}
    \SetKwFunction{GetOptimalParam}{GetOptimalParam}
    \SetKwFunction{DataTransfer}{DataTransfer}
    \SetKwFunction{AdaptiveSampling}{AdaptiveSampling}
    \SetKwFunction{append}{append}
    \SetKwFunction{remove}{remove}
    \SetKwFunction{QueryDB}{QueryDB}
    \SetKwFunction{Sort}{Sort}
    \SetKwFunction{GetSamples}{GetSamples}
	\SetKwInOut{Input}{input}\SetKwInOut{Output}{output}
    \tcp{Source, $EP_{s}$, Destination, $EP_{d}$, Round trip time, $rtt$, Bandwidth, $bw$}
	\Input{Data arguments, $data\_args = \{Dataset , avg\_file\_size, num\_files\}$,
           network arguments, $net\_args = \{EP_s, EP_d, rtt, bw\}$,  
           transfer node arguments, $node\_args = \{num\_nodes,cores, memory, NIC\_speed \}$}
	\Output{Optimal transfer rate, $th_{opt}$}
	\BlankLine
    \SetKwProg{proc}{procedure}{}{}
    \proc{\AdaptiveSampling{$F_s$, $R_s$, $I_s$} }{
    	$D_{s}$ $\leftarrow$ \GetSamples{$Dataset$}  \\
        
    	\ismedian  $\leftarrow$ \Median{$I_s$} \\
        $f_{s,median}$ $\leftarrow$ $F_{s}[e_{s,median}]$ \\
        $\theta_{s,median}$, $th_{hist}$ $\leftarrow$ \GetOptimalParam{$f_{s,median}$} \\
        $th_{cur}$ $\leftarrow$ \DataTransfer{$D_{s,1}$,$p_{s,median}$} \\
        \Log.\append{$net\_args$, $D_{s,i}$,$p_{s,median}$,$th_{cur}$} \\
        $D_{s}$.\remove{$D_{s,1}$} \\
        \For{$D_{s,i}$ \in $D_s$ } {
              \If{$th_{cur}$ $\neq$ $th_{hist}.confidence\_bound$} { 
                   $f_{s,cur}$ $\leftarrow$ \FindClosestSurface{$th_cur$} \\
                   $p_{s,cur}$, $th_{hist}$ $\leftarrow$ \GetOptimalParam{$f_{s,curr}$} \\
                   $th_{cur}$ $\leftarrow$ \DataTransfer{$D_{s,i}$,$p_{s,cur}$} \\
                   \Log.\append{$net\_args$, $D_{s,i}$,$p_{s,cur}$,$th_{cur}$}
              }
        }
    }
    
    $F_s$, $R_s$, $I_s$ $\leftarrow$ \QueryDB{$data\_args$,$net\_args$} \\
	$F{'}_{s}$ $\leftarrow$ \Sort{$F_s$,$I_s$} \\
    \tcp{Set of surfaces, $F_{s}$, Sampling region, $R_{s,k}$, Load intensity, $I_s$ }
	\AdaptiveSampling{$F^{'}_{s}$, $R_{s,k}$, $I_s$}
\caption{Online Sampling}
\label{algo:onlinesampling}
\end{algorithm}\DecMargin{1em}

which is already precomputed during offline analysis and can be found in the sampling region. Achieved throughput value for the transfer is recorded (Lines 2-6). If the achieved throughput is inside the surface confidence bound at point $\theta_s,median$, then the algorithm continues to transfer rest of the data set chunk by chunk. However, if the achieved throughput is outside the confidence bound, that means the current surface is not representing the external load of the network. If achieved throughput is higher than the surface maxima, that means current network load is lighter than the load associated with the surface. Therefore, the algorithm searches the surfaces with lower load intensity tags and find the closest one and perform second sample transfer with parameters of newly found surface maxima. In this way, the algorithm can get rid of half the surfaces at each transfer. At the point of convergence, our algorithm takes the rest of the dataset and starts the transfer process. Changing parameters in real-time is expensive. For example, if a $cc$ value changes from 2 to 4, this algorithm has to open two more server processes and initialize resources. These new processes have to go through TCP's slow start phase as well. Therefore, the algorithm tries to minimize the initial sampling transfers by the adaptive approach. For very large-scale transfers, when data transfer happens for a long period of time, external traffic could change during the transfer. If the algorithm detects such deviation, it uses most recently achieved throughput value to choose the suitable surface and changes the transfer parameters.


\section{Evaluation}
\label{sec:experiments}

\begin{figure*}[t]
    \begin{centering}
\begin{subfigure}[t]{0.32\textwidth}
        \includegraphics[keepaspectratio=true,width=50mm]{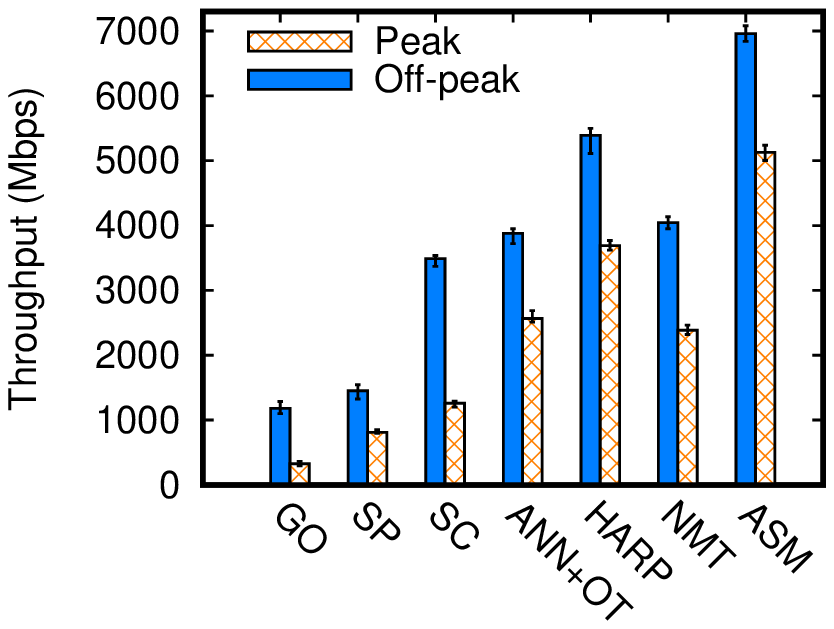}
        \caption{Achieved throughput (small files)}
    \end{subfigure}
    \begin{subfigure}[t]{0.32\textwidth}
        \includegraphics[keepaspectratio=true,width=50mm]{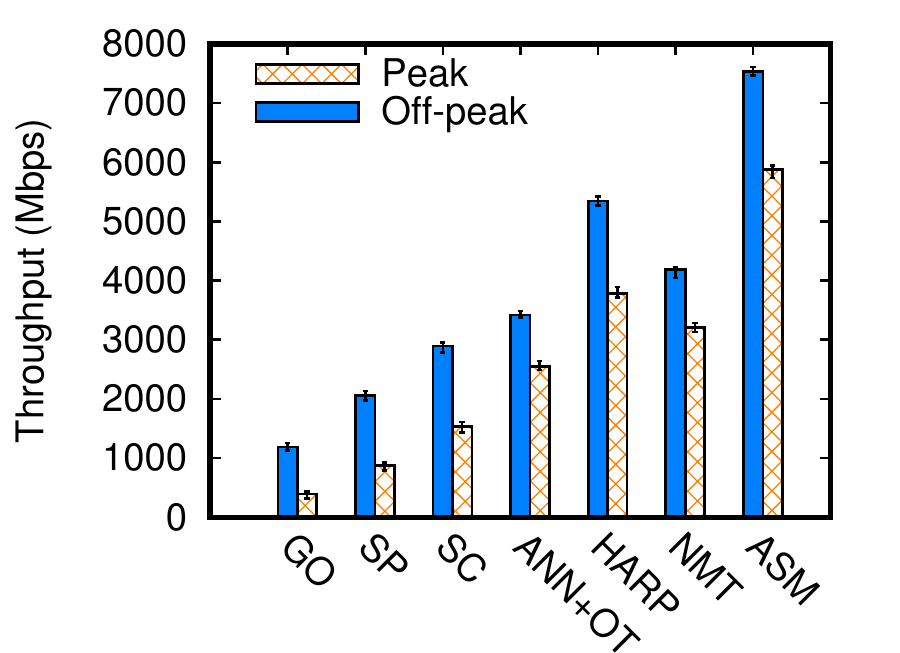}
        \caption{Energy consumption (small files)}
    \end{subfigure}
    \begin{subfigure}[t]{0.32\textwidth}
        \includegraphics[keepaspectratio=true,width=50mm]{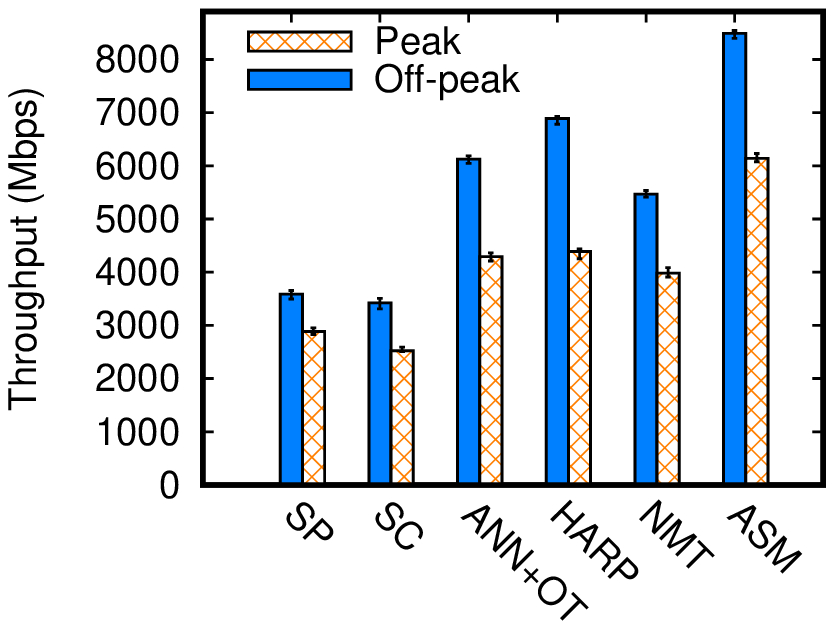}
        \caption{Throughput efficiency (small files)}
    \end{subfigure}
\begin{subfigure}[t]{0.32\textwidth}
        \includegraphics[keepaspectratio=true,width=50mm]{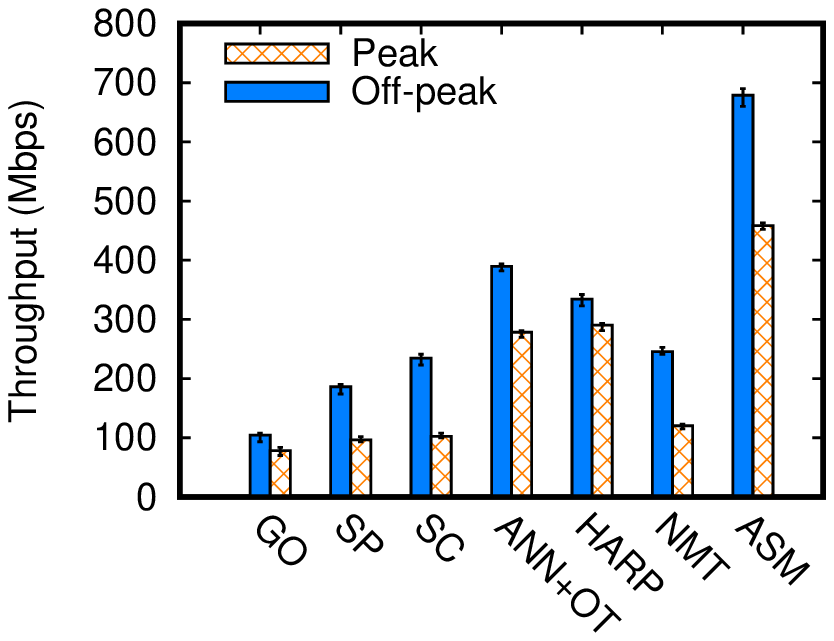}
        \caption{Achieved throughput (medium files)}
    \end{subfigure}
 \begin{subfigure}[t]{0.32\textwidth}
        \includegraphics[keepaspectratio=true,width=50mm]{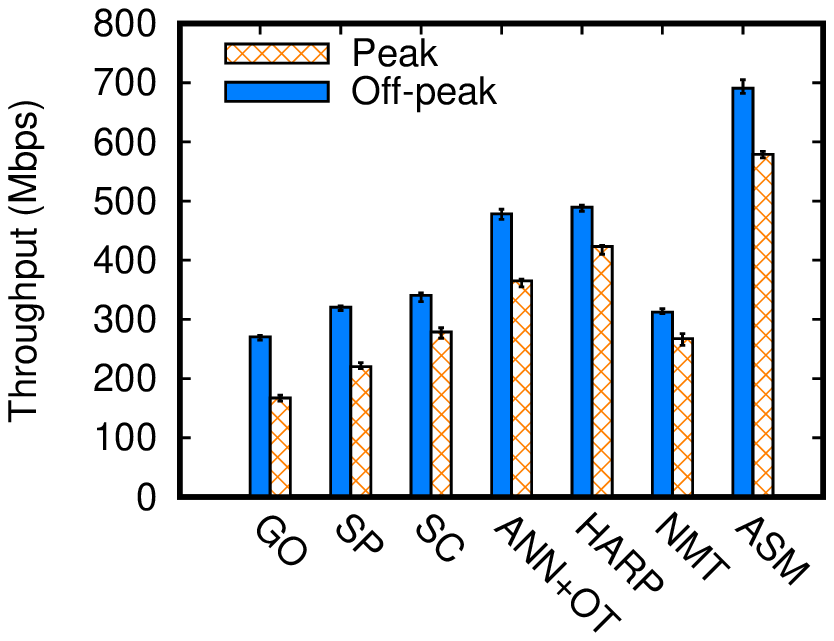}
        \caption{Energy consumption (medium files)}
    \end{subfigure}
    \begin{subfigure}[t]{0.32\textwidth}
        \includegraphics[keepaspectratio=true,width=50mm]{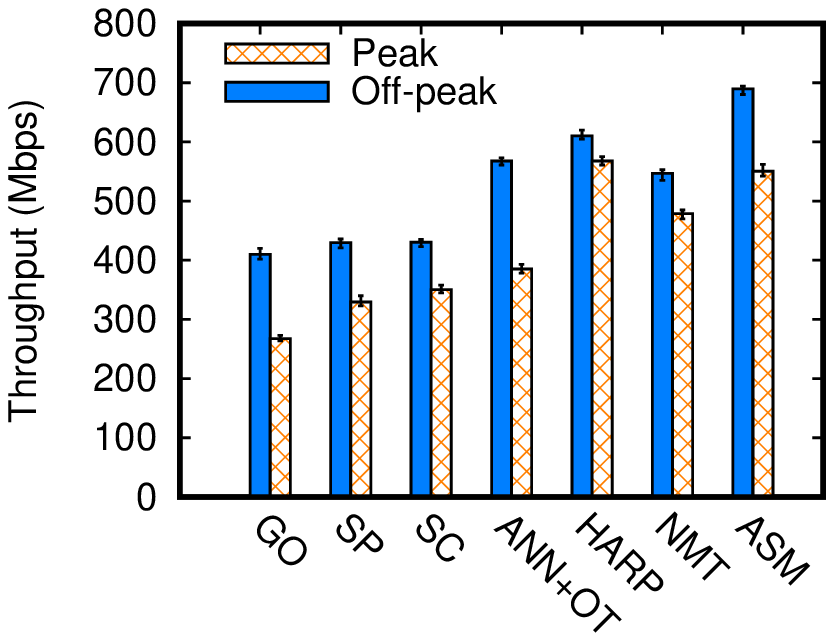}
        \caption{Throughput efficiency (medium files)}
    \end{subfigure}
\begin{subfigure}[t]{0.32\textwidth}
        \includegraphics[keepaspectratio=true,width=50mm]{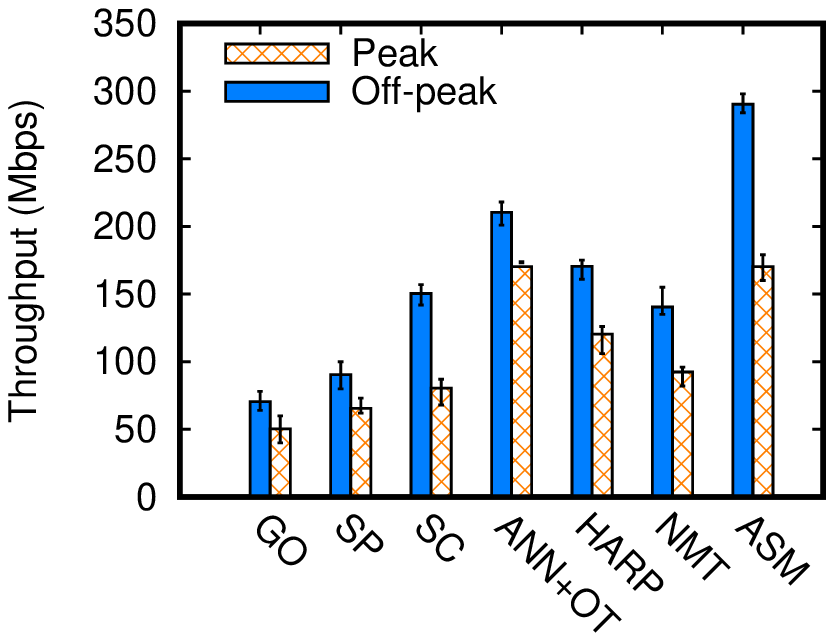}
        \caption{Achieved throughput (large files)}
    \end{subfigure}
    \begin{subfigure}[t]{0.32\textwidth}
        \includegraphics[keepaspectratio=true,width=50mm]{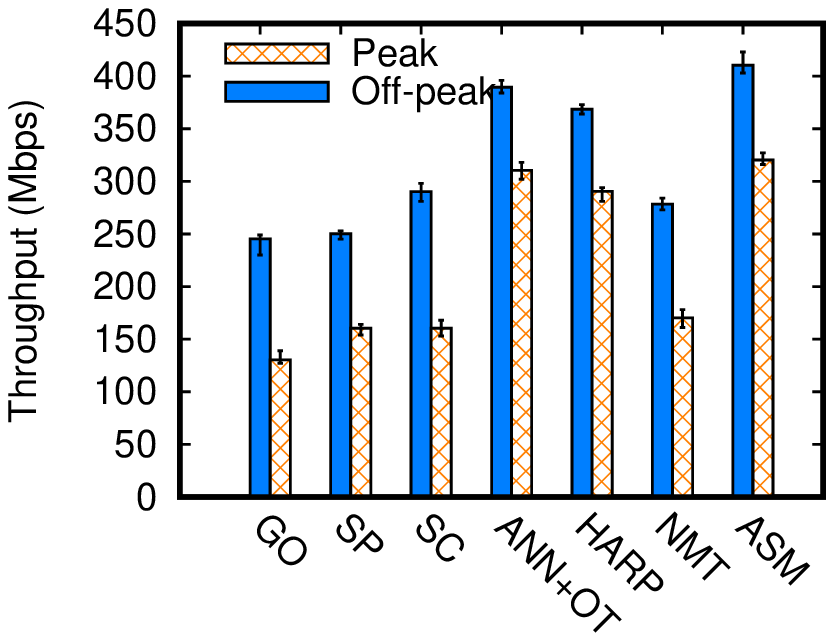}
        \caption{Energy consumption (large files)}
    \end{subfigure}
    \begin{subfigure}[t]{0.32\textwidth}
        \includegraphics[keepaspectratio=true,width=50mm]{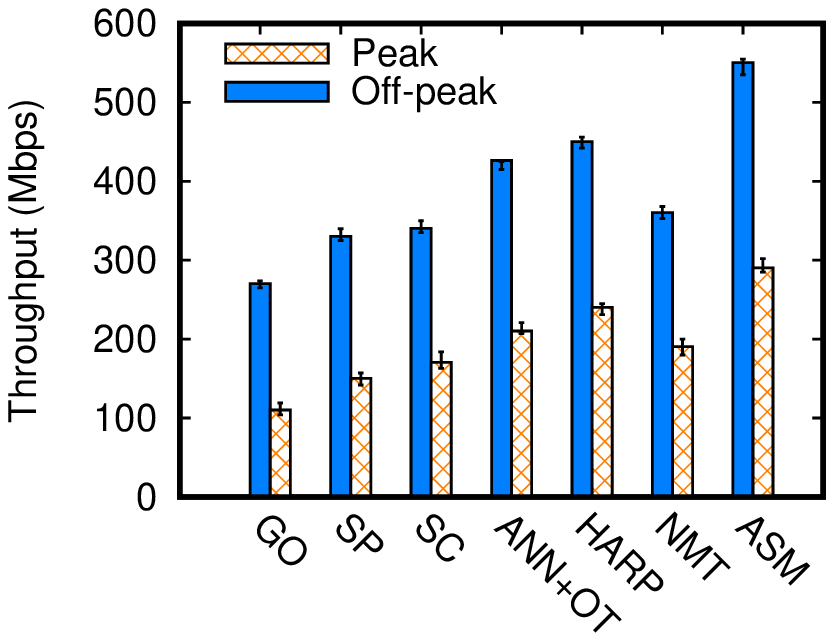}
        \caption{Throughput efficiency (large files)}
    \end{subfigure}
     \caption{Achievable throughput and corresponding energy consumption of different optimization objectives.}
     \vspace{-4mm}
     \label{fig:throughputresults}
     \end{centering}
 \end{figure*}

In the evaluation of our model, we used GridFTP~\cite{globusonline} data transfer logs generated over a six-week period of time. GridFTP is one of the most widely used data transfer protocols in scientific computing, and it is used to transfer 100s of Petabytes of data every year. As the networking environment, we used XSEDE, a collection of high-performance computing resources connected with high-speed WAN and our DIDCLAB testbed. On XSEDE, we performed data transfers between Stampede at Texas Advanced Computing Center (TACC) and Gordon cluster at San Diego Supercomputing Center (SDSC). Table \ref{table:specfications} shows the system and network specifications of our experimental environment.

\begin{table}[tbp]
\centering
\caption{System specification of our experimental environment}
\label{table:specfications}
\begin{tabular}{|l|l|l|c|c|}
\hline
                                                           & \multicolumn{2}{c|}{XSEDE}                              & \multicolumn{2}{c|}{DIDCLAB}                               \\ \hline
                                                           & Stampede                   & Gordon                     & \multicolumn{1}{l|}{WS-10} & \multicolumn{1}{l|}{Evenstar} \\ \hline
\hline
Cores                                                      &                            &                            & 8                          & 4                             \\ \hline
Memory                                                     &                            &                            & 10 GB                      & 4 GB                          \\ \hline
Bandwidth                                                  & \multicolumn{2}{c|}{10 Gbps}                            & \multicolumn{2}{c|}{1 Gbps}                                \\ \hline
RTT                                                        & \multicolumn{2}{c|}{40 ms}                              & \multicolumn{2}{c|}{0.2 ms}                                \\ \hline
\begin{tabular}[c]{@{}l@{}}TCP \\ Buffer size\end{tabular} & \multicolumn{1}{c|}{48 MB} & \multicolumn{1}{c|}{48 MB} & 10 MB                      & 10 MB                         \\ \hline
\begin{tabular}[c]{@{}l@{}}Disk \\ Bandwidth\end{tabular}  & 1200 MB/s                    & 1200 MB/s                    & 90 MB/s                      & 90 MB/s                         \\ \hline
\end{tabular}
\end{table}

We compared our results with the state-of-the art solutions in this area, such as - (1) Static models: Globus (GO) \cite{allen:2012} and Static ANN (SP) \cite{Nine:2015ANN}; (2) Heuristic models: Single Chunk (SC) \cite{engin:dynamicTuning}; (3) Dynamic models: HARP \cite{Engin2016} and ANN+OT \cite{Nine:2015ANN}; and (4) Mathematical models: Nelder-Mead Tuner (NMT) \cite{Ian-bala2016}. Globus uses different static parameter settings for different types of file sizes. SC  also makes parameter decision based on dataset characteristics and network matrices. It asks the user to provide an upper limit for concurrency value. SC does not exceed that limit. HARP uses heuristics to perform a sample transfer. Then the model performs online optimization to get suitable parameters and starts transferring the rest of the dataset. Online optimization is expensive and wasteful as it needs to be performed each time, even for similar transfer requests. ANN+OT  learns the throughput for each transfer request from the historical logs. When a new transfer request comes, model asks the machine learning module for suitable parameters to perform first sample transfer. Then it uses recent transfer history to model the current load and tune the parameters accordingly. The model only relies on historical data and always tends to choose the local maxima from historical log rather than the global one.  Nelder-Mead Tuner implements a direct search optimization which does not consider any historical analysis, rather tries to reach optimal point using reflection and expansion operation. We tested those models three different networks: (1) between two XSEDE nodes; (2) between two DIDCLab nodes; and (3) between DIDCLab and XSEDE nodes.  

\begin{figure}[t]
\begin{centering}
\includegraphics[keepaspectratio=true,angle=0,width=80mm]{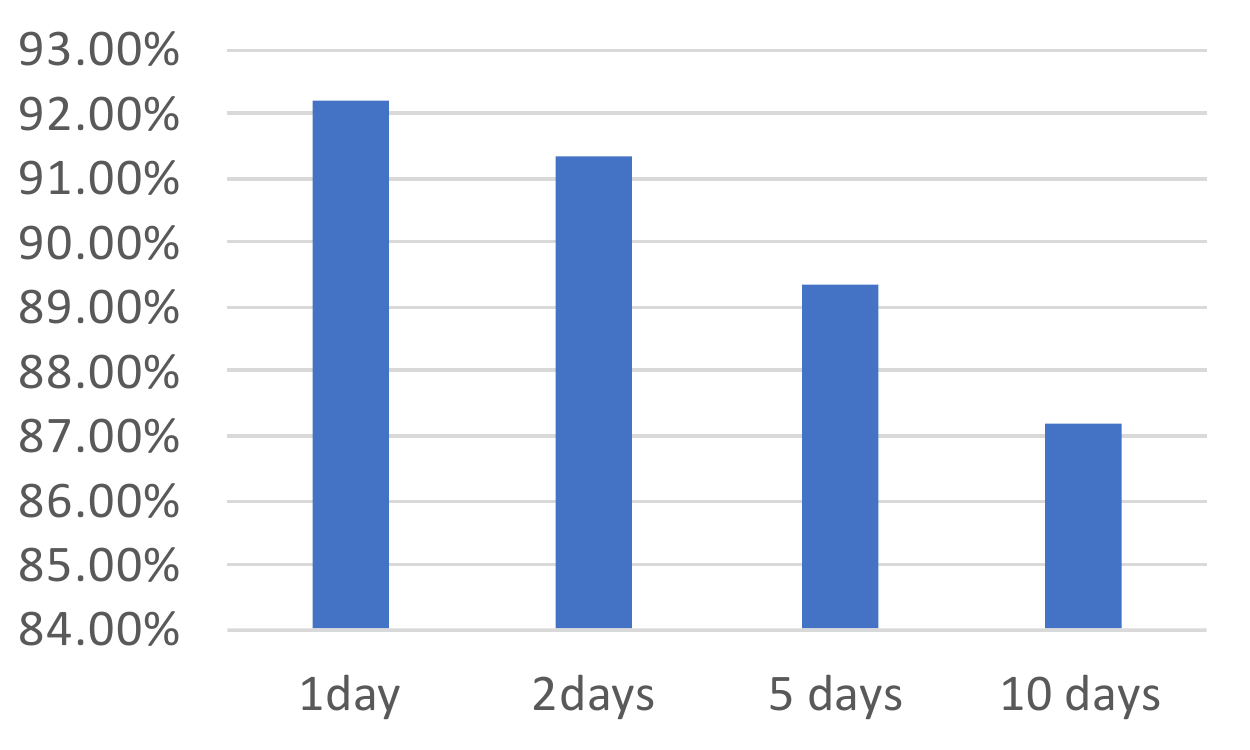}
	\end{centering}
    \vspace{-3mm}
\caption{ Model accuracy over periodic offline analysis.} \label{fig:periodicaccuracy}
\end{figure}

\subsection{Performance analysis}
We tested our model with data transfer requests those are completely different from the historical logs used in the model. To ensure that we computed the list of all unique transfers and split the list as 70\% for training the model and 30\% for test purpose. We also evaluated our model on both peak and off-peak hours to measure performance under different external load conditions. Achievable throughput is highly dependent on the average file size of the dataset. 

In order to evaluate the accuracy of our model for different types of average file sizes, we partitioned transfer requests into three groups - small, medium and large. Then we compared average achievable throughput so that we can evaluate the model in a more fine-grained way. Figure \ref{fig:throughputresults} shows the comparison of our proposed Adaptive Sampling Module (ASM) with the other state-of-the-art solutions mentioned above.
In all three networks and for all datasets, ASM outperforms all other models. The second best performing model in all of these experiments is HARP~\cite{Engin2016}. In the XSEDE to XSEDE experiments (Figure~\ref{fig:throughputresults}(a-c)) ASM outperforms HARP by 29\% for small datasets, 40\% for medium datasets, and 23\% for large datasets. Adaptive sampling solves the slow convergence problem with the more accurate pre-constructed representation of throughput surfaces. Our model also gets rid off all the surface regions those proved suboptimal for different background traffic. Moreover, it has a fast online module with adaptive sampling that can converge faster and reduces the suboptimal convergence time. Moreover, our model obtains more impressive performance during peak hours. It outperforms HARP by 38\%, 55\%, and 39\% for small, medium, and large datasets respectively. Peak hour periods are challenging to model, and the result shows that our offline analysis is resilient enough to achieve better results in such network environment, with the help of adaptive sampling module. 

Figure \ref{fig:throughputresults}(d-f) shows the performance of different models in our DIDCLAB testbed. 
Again, our model (ASM) outperforms all the existing models. It achieves 100\% performance improvement over HARP during small file transfers during off-peak hours. It outperforms HARP by 41\% during medium dataset transfers. However, for large files, the performance improvement is only 13\% and during peak hours HARP actually does slightly better than our model. HARP's performance basically depends on its regression accuracy in this case. 


In Figure \ref{fig:throughputresults} (g-i), we report the performance of these models between DIDCLAB to XSEDE network. This is a quite busy Internet connection which makes it more challenging. 
In this network too our model performed better than all the mentioned models. For small dataset, our model outperforms its closest competitor ANN+OT by 38\%. It outperforms HARP by 22\% during large dataset transfers.
Our online module needs almost constant time to agree on the parameters. Among the existing models that we have tested so far, only HARP uses the online optimization which could be expensive, however, rest of the models can perform transfers in constant time. 

\subsection{Performance of Offline analysis}
Among the above-mentioned models, static, heuristics, and mathematical optimization models do not require any historical analysis, however, our model requires extra historical analysis. Therefore, a natural question would be, how often do we have to perform the offline analysis? The answer is, we do not need to perform offline analysis before every single data transfer request, rather it can be done periodically. Figure \ref{fig:periodicaccuracy} shows the impact of offline analysis frequency on the accuracy of the model. Offline analysis performed once a day is enough to reach 92\% accuracy. Model accuracy decreases slightly to 87\% even for cases where offline analysis is performed once in 10 days. This shows the model could converge faster, even when offline analysis are performed 10 days apart.

\subsection{Performance of Dynamic Tuning}
Adaptive Sampling Module(ASM) performs online sampling and uses the network information to query the offline analysis for optimal parameters along with the achievable throughput, $T_{predict}$. The optimal parameters are used for the next sample transfer. Then we measure the actual throughput achieved, $T_{achieved}$. As our model converges $T_{achieved}$ gradually, it gets closer to the $T_predict$. To measure the accuracy of the model we used the following metric:

\begin{figure}[t]
\begin{centering}
\includegraphics[keepaspectratio=true,angle=0,width=90mm]{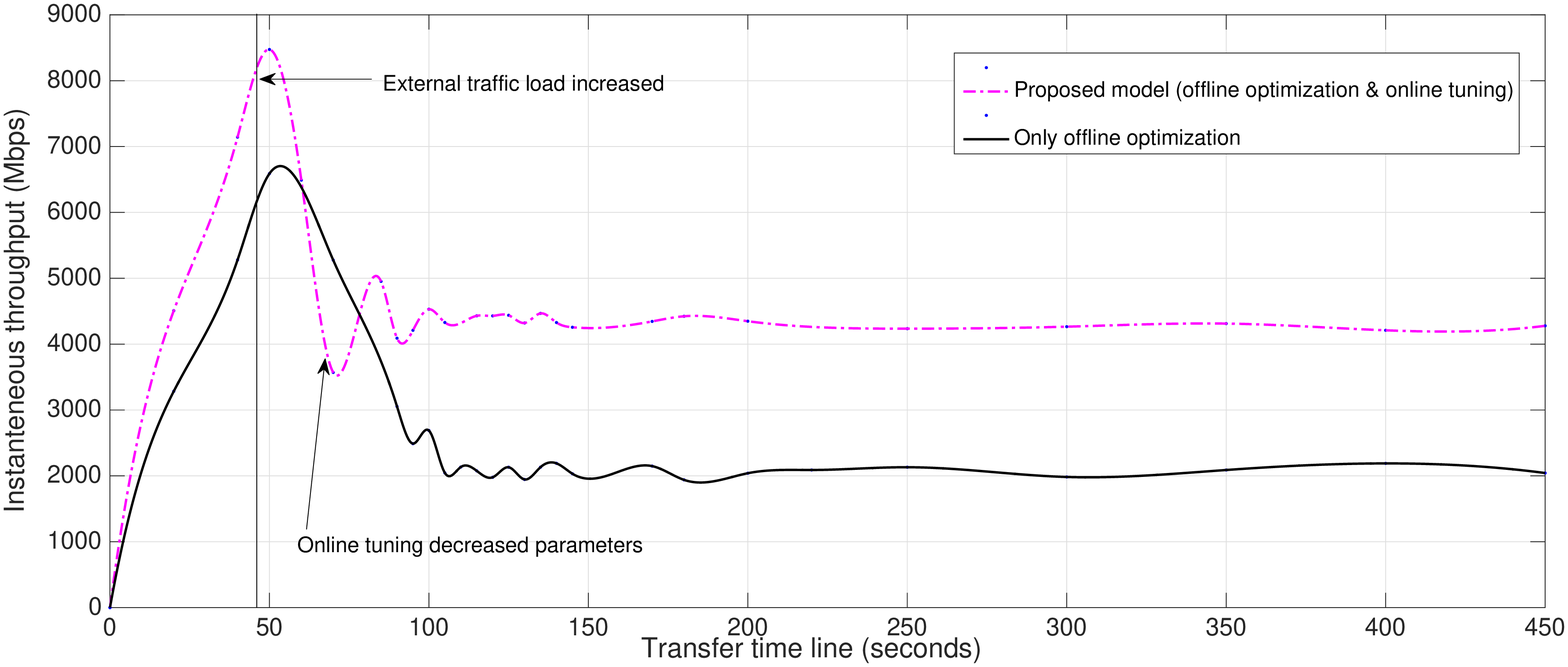}
	\end{centering}
    \vspace{-3mm}
\caption{ Convergence of DT model} \label{fig:multiuser-case}
\vspace{-5mm}
\end{figure}

\begin{equation}
\mathrm{Accuracy} = \dfrac{|T_{achieved }- T_{predict}|}{T_{predict}} \times 100
\end{equation}

\begin{figure}[t]
\begin{centering}
\includegraphics[keepaspectratio=true,angle=0,width=80mm]{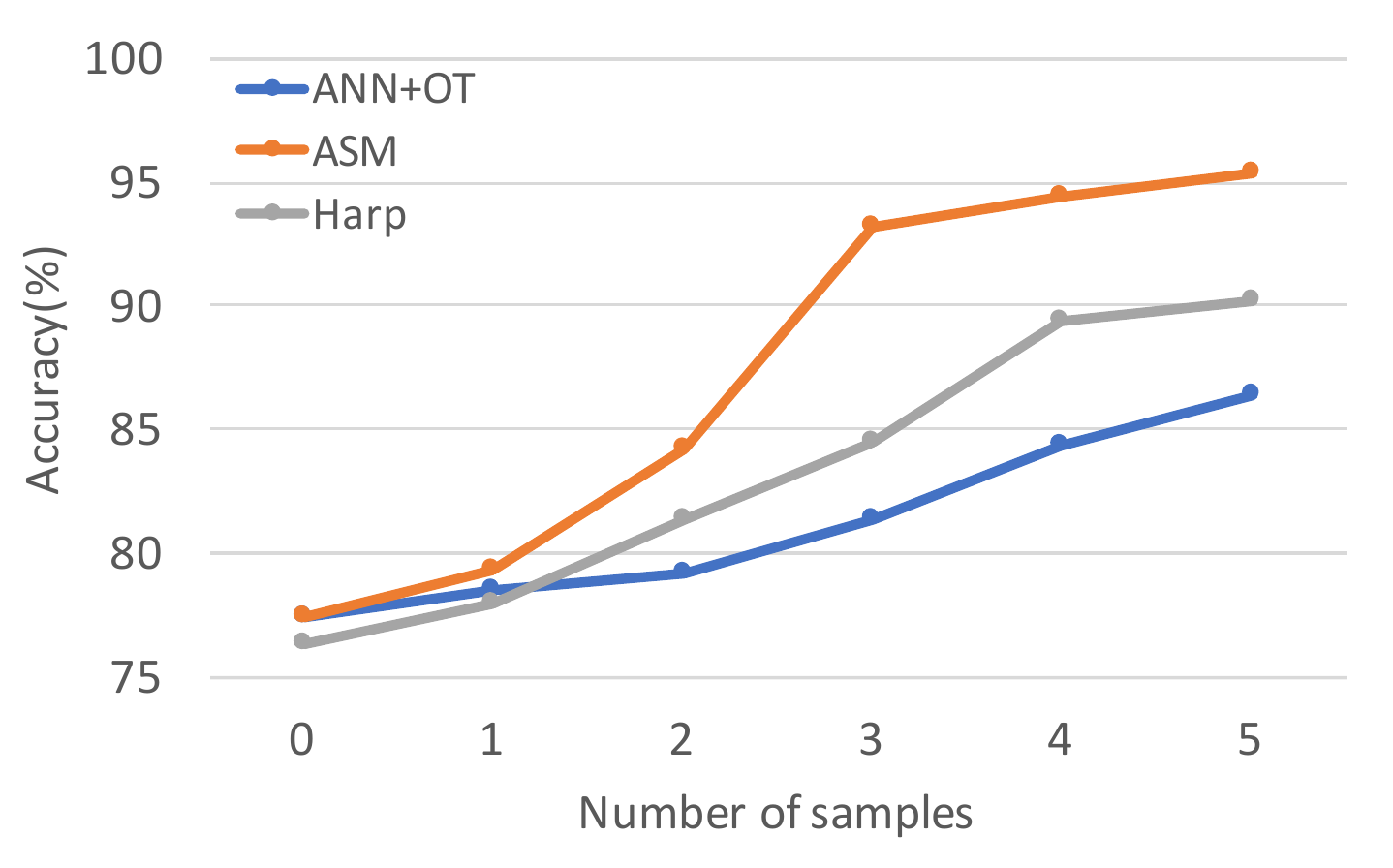}
	\end{centering}
    \vspace{-3mm}
\caption{Prediction accuracy of different models with respect to number of sample transfers (those uses online sampling). } 
\label{fig:predictionaccuracy}
\end{figure}

Figure \ref{fig:predictionaccuracy} shows a comparison of the accuracy of throughput prediction models. HARP can reach up to 85\% with 3 sample transfers along with high online computation overhead. ANN+OT can reach 87.32\% accuracy. Our model achieves almost 93\% accuracy with three sample transfers for any types of dataset and then it saturates. It shows that our offline cubic spline interpolation can model the network more accurately and adaptive sampling can ensure faster convergence towards the optimal solution.

\subsection{Fairness analysis}

\begin{figure}[t]
\begin{centering}
\includegraphics[keepaspectratio=true,angle=0,width=88mm]{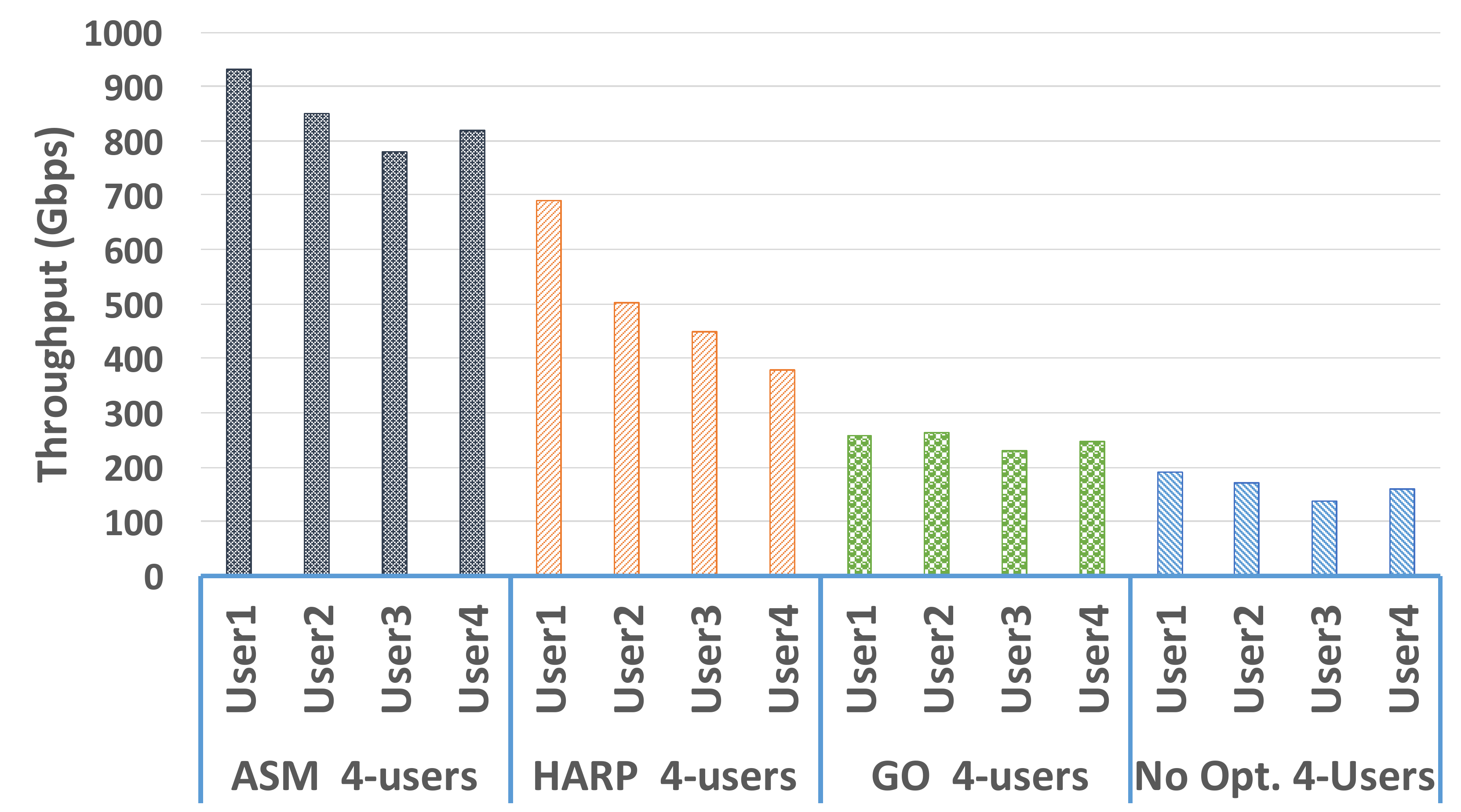}
	\end{centering}
    \vspace{-3mm}
\caption{ Achievable throughput (Gbps) for different models in a multi-user scenario in Chameleon Cloud between a CHI-UC and a TACC node.} \label{fig:multiuser-case}
\vspace{-5mm}
\end{figure}

\begin{figure}[t]
\begin{centering}
\includegraphics[keepaspectratio=true,angle=0,width=88mm]{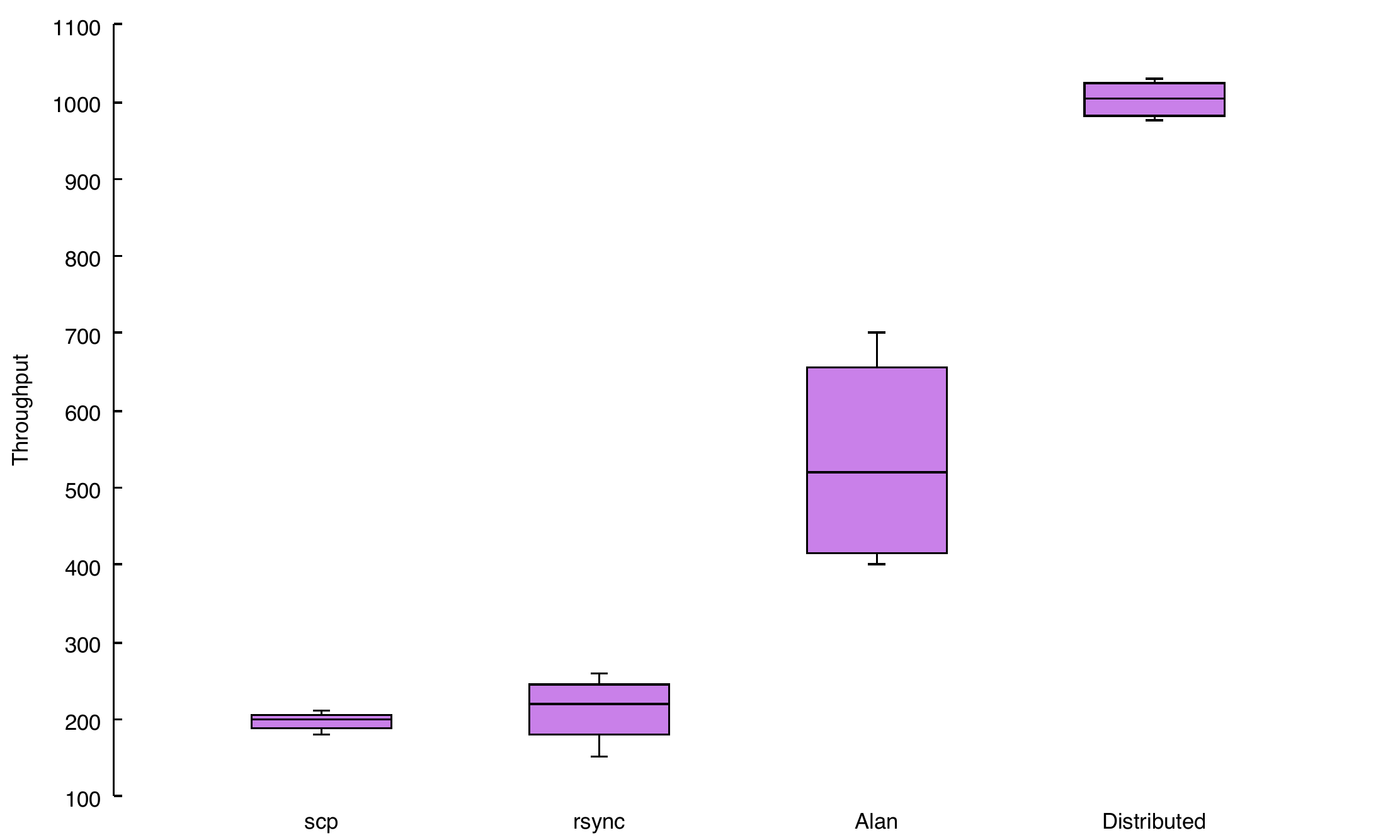}
	\end{centering}
    \vspace{-3mm}
\caption{ Achievable throughput (Gbps) for different models in a multi-user scenario in Chameleon Cloud between a CHI-UC and a TACC node.} \label{fig:multiuser-case}
\vspace{-5mm}
\end{figure}

One potential question would be: ``What will happen if multiple users try to use the same optimization technique to improve their transfer throughput? Would they hurt each other's performance and suffer a performance degradation rather than improvement?'' Figure \ref{fig:multiuser-case} shows the performance of these models under multi-user scenario. We used CHI-UC and TACC nodes in Chameleon cloud to test the performance when multiple users are transferring data simultaneously using same optimization technique. The figure shows four such cases: (1) ASM with 4-users; (2) HARP with 4-users; (3) GO with 4-users; and (4) No optimization with 4-users. In No Optimization case, users use same static parameter setting $(p = pp = cc = 1)$. Among many different models, we choose the closest competitor HARP and two baseline models for performance comparison. ASM outperforms its closest competitor HARP in the multi-user scenario by 1.7x, GO by 3.4x, and the default (no optimization) case by 5x.  This shows that ASM can utilize the available network bandwidth much better compared to the other models.

Another question would be whether fairness among different users is preserved or not, since fairness is an important feature when multiple users are using a shared network.As seen in the figure, ASM preserves fairness among users by maintaining a low standard deviation of achievable throughput by different users. For ASM, the standard deviation is only 54.98, whereas this value is almost double for HARP (115.49). HARP performs real-time sampling only at the beginning and can aggressively set the parameters which might hurt the throughput of other users. However, ASM periodically checks for network status and the adaptive sampling method can intelligently adjust the best parameter value when the external load changes dynamically. When multiple users are using ASM in a shared network, everyone tries to aggressively set the parameters until individual ASM instances can detect performance drop and starts recalculating the parameters. Eventually, they can adjust their parameters to get a fair share of the available throughput. HARP does not have this ability as it sets the parameters at the beginning. The user who starts initial probing first can aggressively set the parameters and might slightly gain advantage over the other users. That is why we can see a gradual decrease to the subsequent users who perform probing later. GO and No Optimization cases also provide a fair share of throughput because all the users are using same static parameter setting, however, their achievable throughput is way less than ASM. 


\section{Related Work}
\label{sec:Related Work}
Earlier work on application level tuning of transfer parameters mostly proposed static or non-scalable solutions to the problem with some predefined values for some generic cases~\cite{globusonline, R_Hacker02, R_Crowcroft98, R_Dinda05, JGrid_2012, ICSIS_2009, Thesis_2005, Royal_2011}. The main problem with such solutions is that they do not consider the dynamic nature of the network links and the background traffic in the intermediate nodes. 

Yin et al. ~\cite{R_Yin11} proposed a full second order model with at least three real-time sample transfers to find optimal parallelism level. The relationship between parallel streams and throughput along with other parameters are more complex than second order polynomials. Moreover, it does not provide concurrency and pipelining. 
 Yildirim et al.~\cite{balance} proposed PCP algorithm which clusters the data based on file size and performs sample transfers for each cluster. Sampling overhead could be very high in this model as it does not consider any historical knowledge for optimization. 

Engin et al. ~\cite{Engin2016} proposed HARP which uses heuristics to provide initial transfer parameters to collect data about sample transfers. After that model performs the optimization on the fly where it has to perform cosine similarity over the whole dataset which might prove expensive. Even if the optimization and transfer task can be parallelized, it could be wasteful as the same optimization needs to be performed for similar transfers every time a similar transfer request is made.  
 
 Prasanna et al. ~\cite{Ian-bala2016} proposed direct search optimization that tune parameters on the fly based on measured throughput for each transferred chunk. However, it is hard to prove the convergence and sometimes hard to predict the rate of convergence. Some cases, it requires 16-20 epochs to converge which could lead to under-utilization. 
 
 Different from the existing work, we address the following issues in this paper:
 {\em (i)} Lower order regression model can underfit the data when higher order polynomials can introduce overfitting, in addition, to compute cost and sampling overhead. For small to moderate size of data transfer requests, slow convergence could lead to severe under-utilization.
 %
 {\em (ii)} Model free dynamic approaches suffer from convergence issue. And convergence time depends on the location of initial search point.
 %
 {\em (iii)} Searching parameters during the transfer could introduce many overheads. Opening a TCP connection in the middle of the transfer introduces a delay due to slow start phase. When initial parameters are far away from optimal solution slow convergence could lead to under-utilization of the network bandwidth which could hurt the overall bandwidth.  
 %
 {\em (iv)} Optimization based on historical log should not be done during the transfer, offline analysis can reduce the real-time computing overhead. 

\section{Conclusion}
\label{sec:conclusion}
In this study, we have explored a novel big data transfer throughput optimization model that relies upon offline mathematical modeling and online adaptive sampling. Existing literature contains different types of throughput optimization models that range from static parameter based systems to dynamic probing based solutions. Our model eliminates online optimization cost by performing the offline analysis which can be done periodically. It also provides accurate modeling of throughput which helps the online phase to reach near optimal solution very quickly. For large scale transfers when external background traffic can change during transfer, our model can detect the harsh changes and can act accordingly. Adaptive sampling module can converge faster than existing solutions. The overall model is resilient to harsh network traffic changes. We performed extensive experimentations and compared our results with best known existing solutions. 
Our model outperforms its closest competitor by 1.7x and the default case by 5x in terms of the achieved throughput. 
It also converges faster, and achieves up to 93\% accuracy compared with the optimal achievable throughput possible on the tested networks.

As future work, we are planning to increase the achievable throughput further by reducing the impact of TCP slow start phase. Another interesting path is to reduce the overhead introduced by real-time parameter changes. We are also planning to investigate other application-layer protocol parameter sets that can be optimized to achieve even better performance.

\bibliographystyle{IEEEtran}
\bibliography{main,didc,misc,rela_work,sc2017}

\end{document}